\documentclass[review]{elsarticle}

\usepackage{lineno,hyperref}
\usepackage{xspace}
\modulolinenumbers[200]
\usepackage{bm}
\usepackage{amssymb}
\usepackage{hyperref}
\usepackage{amsmath}
\usepackage[dvipsnames]{xcolor}
\usepackage{soul}

\newcommand{\pt}{\ensuremath{p_{\rm{T}}}\xspace}
\newcommand{\raa}{\ensuremath{R_{\rm{AA}}}\xspace}

\journal{Journal of \LaTeX\ Templates}

\begin{document}

\begin{frontmatter}

\title{Intriguing similarities between high-$p_{\rm T}$ particle production in pp and A-A collisions}

\author{Aditya Nath Mishra, Antonio Ortiz and Guy Pai{\'c}}
\address{Instituto de Ciencias Nucleares, Universidad Nacional Aut\'onoma de M\'exico, \\ Apartado Postal 70-543,
M\'exico Distrito Federal 04510, M\'exico}





\begin{abstract}

In this paper we study the particle production at high transverse momentum ($\pt>8$\,GeV/$c$) in both pp and Pb-Pb collisions at LHC energies. The characterization of the spectra is done using a power-law function and the resulting power-law exponent ($n$) is studied as a function of $x_{\rm T}$ for minimum-bias pp collisions at different $\sqrt{s}$. The functional form of $n$ as a function of $x_{\rm T}$ exhibits an approximate universal behavior. PYTHIA~8.212 reproduces the scaling properties and therefore, it is used to study the multiplicity-dependent particle production. Going from low to high multiplicities, the power-law exponent decreases. A similar behavior is also observed in heavy-ion collisions when one studies the centrality-dependent particle production. The interpretation of heavy-ion results requires the quantification of the impact of this correlation (multiplicity and high $p_{\rm T}$) on jet-quenching observables.
\end{abstract}

\begin{keyword}
Hard scattering \sep Jets and heavy flavor physics \sep Relativistic heavy-ion collisions
\end{keyword}

\end{frontmatter}

\linenumbers

\section{Introduction}

The similarity between analogous observables in large (A-A) and small (pp and p-A) collisions systems has been extensively studied by the heavy-ion community~\cite{edward:2010aip,edward:2010epjc,edward:2014epjc,edward:2016prd,edward:2016Rprd}.  A vast number of quantities as a function of the charged-particle density (${\rm d}N_{\rm ch}/{\rm d}\eta$) in small systems have been documented in recent works~\cite{Loizides:2016tew}.  These observables (azimuthal anisotropies, radial flow and strangeness enhancement~\cite{Abelev:2013haa,Khachatryan:2016txc,ALICE:2017jyt}) have been measured in the low- and intermediate-transverse momentum regimes ($\pt<8$\,GeV/$c$).  For higher transverse momenta, the traditional treatments intend to isolate the QGP effects using reference data where the formation of a medium is not expected. Minimum-bias proton-proton collisions have been used for this purpose. However, now this assumption is questionable~\cite{Zakharov:2013gya,Mangano:2017plv}.

The PHENIX collaboration has collected data of nucleus-nucleus collisions from $\sqrt{s_{\rm NN}}=62.4$ up to 200\,GeV, and the results were compared with those from Pb-Pb collisions at $\sqrt{s_{\rm NN}}=2.76$\,TeV. Using the so-called fractional momentum loss, particle production at high \pt ($\pt>8$\,GeV/$c$) in A-A collisions was compared with the one in minimum-bias pp collisions at the corresponding center-of-mass energy.  Surprisingly, this quantity was found to scale better with $\langle {\rm d}N_{\rm ch}/{\rm d}\eta \rangle$ and with the Bjorken energy density times the equilibration time ($\epsilon_{\rm Bj}\tau_{0}$) than with the number of participants obtained using the Glauber model~\cite{Adare:2015cua}. These results motivated further studies which confirmed the scaling even at the top LHC energy of $\sqrt{s_{\rm NN}}=5.02$\,TeV~\cite{Ortiz:2017cul}. Similarly, recent results of the ALICE collaboration show that the nuclear modification factors (\raa) in Pb-Pb collisions at  $\sqrt{s_{\rm NN}}=2.76$\,TeV and 5.02\,TeV~\cite{Acharya:2018qsh} and Xe-Xe collisions at $\sqrt{s_{\rm NN}}=5.44$\,TeV~\cite{Acharya:2018eaq} scale with $\langle {\rm d}N_{\rm ch}/{\rm d}\eta \rangle$. This suggests that multiplicity (or energy density) may play an important role to describe the high-\pt particle production in heavy-nuclei collisions.

The correlation between particle production at high transverse momentum and the large underlying event activity has been extensively documented for pp and p-Pb collisions~\cite{ALICE:2011ac,Ortiz:2017jaz,Armesto:2015kwa,Alvioli:2014eda,Ortiz:2018vgc}. Namely, for small systems the underlying event activity increases with increasing the leading particle transverse momentum. The production of high-momentum particles in Pb-Pb systems could also bias towards high-multiplicity nucleon-nucleon collisions. Therefore it is important to perform a systematic study of the system-size dependence of particle production at high \pt.  Moreover, the study of the transverse momentum spectra in a large momentum range is a very good laboratory to observe the successive dominance of the gluon and quark contributions~\cite{Horowitz:2011gd}.


In the present work we do a comprehensive study of the multiplicity dependence of particle production at high transverse momentum ($\pt>8$\,GeV/$c$) in pp collisions at LHC energies. The results are then compared with LHC A-A data. Although this kind of studies is important to understand the propagation of a hard probe within the medium, they have not been reported so far. The message of the present paper is that the {\bf shape} of \raa for high-\pt particles is not fully attributed to the parton energy loss; since as we will demonstrate, a similar shape is observed for the analogous ratios in pp collisions, i.e., high-multiplicity \pt spectra normalized to that for minimum-bias events.

The paper is organized as follows: Sec. 2 describes how the high-\pt production is characterized in terms of a power-law function as well as the description of the data which were used in this analysis. The results and discussions are displayed in Sec. 3 and final remarks are presented in Sec. 4.

\section{Particle production at large transverse momenta}

In heavy-ion collisions particle production at high \pt is commonly used to study the opacity of the medium to the jets. Experimentally, the medium effects are extracted by means of the nuclear modification factor, \raa, which is defined as:
\begin{equation}
	R_{\rm AA}=\frac{{\rm d}^2N_{\rm AA}/{\rm d}y{\rm d}\pt}{\langle N_{\rm coll} \rangle{\rm d}^2N_{\rm pp}/{\rm d}y{\rm d}\pt}
\end{equation} 
where ${\rm d}^2N_{\rm AA}/{\rm d}y{\rm d}\pt$  and ${\rm d}^2N_{\rm pp}/{\rm d}y{\rm d}\pt$ are the invariant yields measured in A-A and minimum-bias pp collisions, respectively.  The ratio is scaled by the average number of binary nucleon-nucleon collisions ($ N_{\rm coll} $) occurring within the same A-A interaction, which is usually obtained using Glauber simulations~\cite{Alver:2008aq,Loizides:2014vua}.  The resulting ratio is supposed to account  (at least from 8 GeV/$c$ onward) for the so-called jet quenching whereby the high-momentum partons would  be ``quenched'' in the hot system created in the nuclei collisions.

The definition of \raa involves two important aspects:
\begin{enumerate}
	\item The absolute normalization of the \pt spectra in minimum-bias pp collisions obtained from the Glauber model~\cite{Alver:2008aq,Loizides:2014vua}, means to represent  the average number of minimum-bias pp collisions (binary collisions) that the colliding nucleons have suffered within the same heavy-ion collision. 
	\item The shape of \raa at high \pt is determined by the different probability for the occurrence of a hard scattering which is larger in heavy-ion collisions than in pp collisions and is proportional to the path length in the medium and its characteristic transport coefficient $\hat{q}$.
\end{enumerate} 

The rationale for the procedure is the following. The normalized ratio should give us the probability that partons normally produced in the multiple binary pp collision get degraded in the hot and dense system created in the heavy-ion collision, resulting in a suppression of the ratio ($\raa<1$).  For instance, in the 0--5\% Pb-Pb collisions at the LHC energies the suppression is about 7--8 for \pt of around 6--7\,GeV/$c$~\cite{Khachatryan:2016odn, Acharya:2018qsh}.  For higher \pt, \raa  exhibits a continuous rise and  approaches unity~\cite{Khachatryan:2016odn}. As suggested by the \pt-differential baryon-to-meson ratio~\cite{Abelev:2014laa,Adam:2015kca}, for \pt larger than 8\,GeV/$c$ radial flow effects are negligible and therefore, the shape of \raa is expected to be dominated by parton energy loss.

\begin{table}[h!]
	\centering
	\begin{tabular}{ |p{1.8cm} p{1.2cm} p{1.2cm} p{1.2cm}  p{1.2cm}p{1.2cm}|  }
		\hline
		Class name & I & II   &  III   &  IV   &  V   \\
		\hline
		$N_{\rm ch}$ & 0 -- 5 & 6 -- 10 & 11 -- 15  & 16 -- 20 &  21 -- 25 \\
		fraction &  10.45\%   &    15.68\%   &   14.79\% &  13.78\%  &  12.34\%  \\
		\hline
		\hline
		Class name & VI & VII   &  VIII   &  IX   &  X   \\
		\hline  
		$N_{\rm ch}$ & 26 -- 30   & 31 -- 35  & 36 -- 40 & 41 -- 50 & $\geq51$  \\
		fraction &   10.39\%   &  8.08\%    &  5.78\%   &   6.09\%    &   2.61\%   \\ 
		\hline
	\end{tabular}
	\caption{Event multiplicity classes based on the number of charged particles ($N_{\rm ch}$) within $|\eta|<0.8$. Results are presented for pp collisions at $\sqrt{s}=13$\,TeV simulated with PYTHIA~8.212 (tune Monash~2013). The contributions to the inelastic cross section (fraction) are also displayed.}
	\label{tab:1}
\end{table}

The underlying assumption to that paradigm is that the pp spectrum does not have a marked dependence on event multiplicity. However, this is not true as indicated by the sphericity analysis as a function of charged-particle multiplicity~\cite{Abelev:2012sk}. The results indicate that even at high multiplicity the abundance of jetty-like events is not negligible, although its contribution is overestimated by the QCD-inspired Monte Carlo generators like PYTHIA~8.212~\cite{CMS:2016gme,Sjostrand:2014zea}. Based on models,  in pp collisions the jet contribution increases with increasing multiplicity~\cite{Ortiz:2016kpz}, this effect contributes to the increase of the particle production at high transverse momentum.

In the present paper we study the shape of the \pt spectra of charged particles measured in heavy-ion and pp collisions separately. The aim is to discuss the origin of the rise  of the \raa for $\pt>6$\,GeV/$c$. Since PYTHIA~8.212 (tune Monash 2013~\cite{Skands:2014pea}) reproduces rather well many features of LHC data~\cite{Adam:2015pza,Ortiz:2013yxa}, we base our studies on PYTHIA~8.212 simulations of pp collisions for different multiplicity classes. The multiplicity classes are defined based on the number of primary charged particles within $|\eta|<0.8$. Table~\ref{tab:1} shows the different event classes and their corresponding contributions to the inelastic cross section for pp collisions at $\sqrt{s}=13$\,TeV.  

\begin{figure}[t!]
	\begin{center}
		\includegraphics[keepaspectratio, width=0.70\columnwidth]{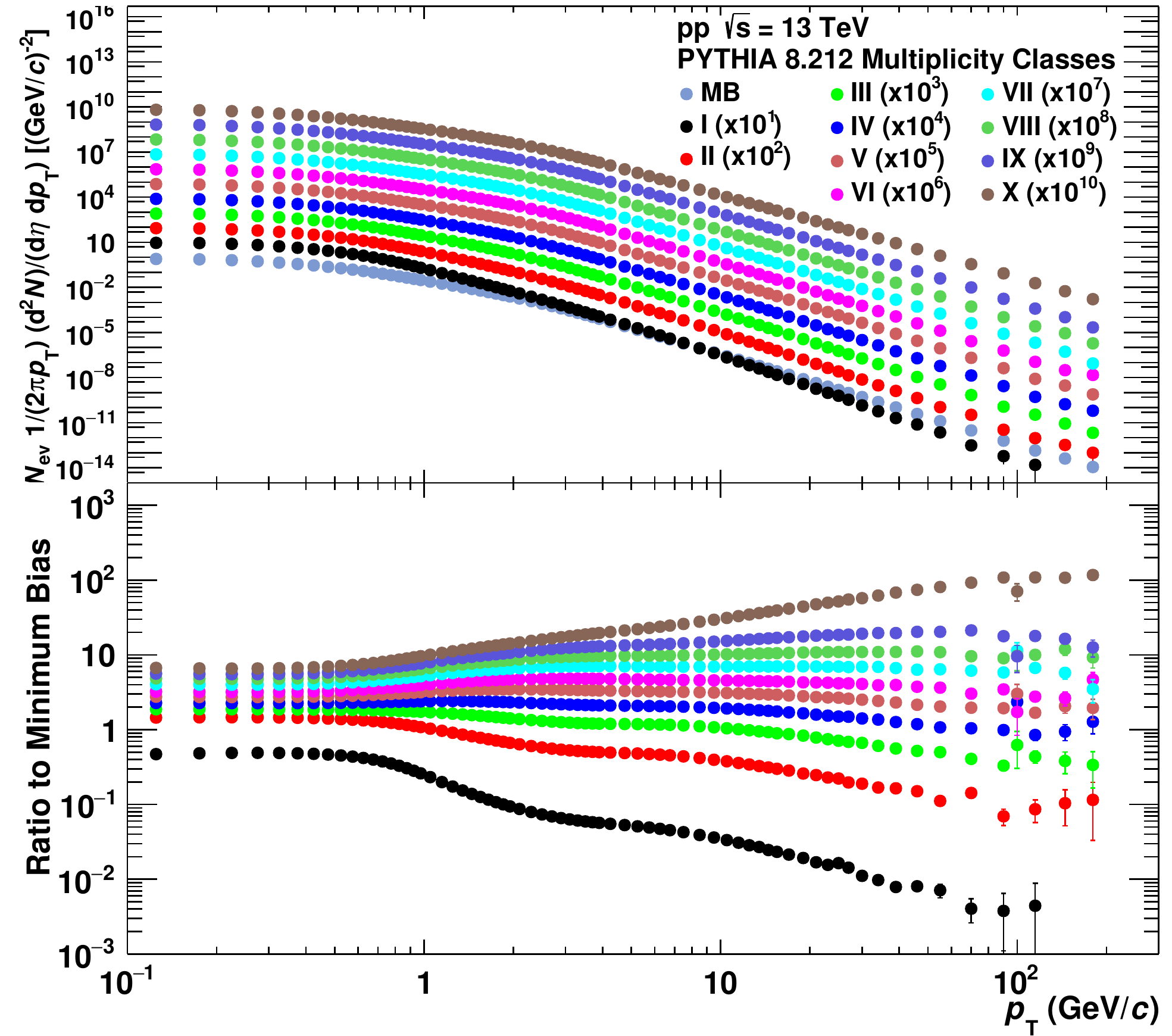}
		\caption{\label{fig:1} Transverse momentum distributions of charged particles for different multiplicity classes in pp collisions at $\sqrt{s}=13$\,TeV simulated with PYTHIA 8.212. The ratios of the multiplicity-dependent \pt spectra to minimum-bias (MB) \pt spectrum are shown in the bottom panel. The spectra are scaled by different factors to improve the visibility.} 
	\end{center}
\end{figure}

\begin{figure*}[t!]
	\begin{center}
		\includegraphics[keepaspectratio, width=1.05\columnwidth]{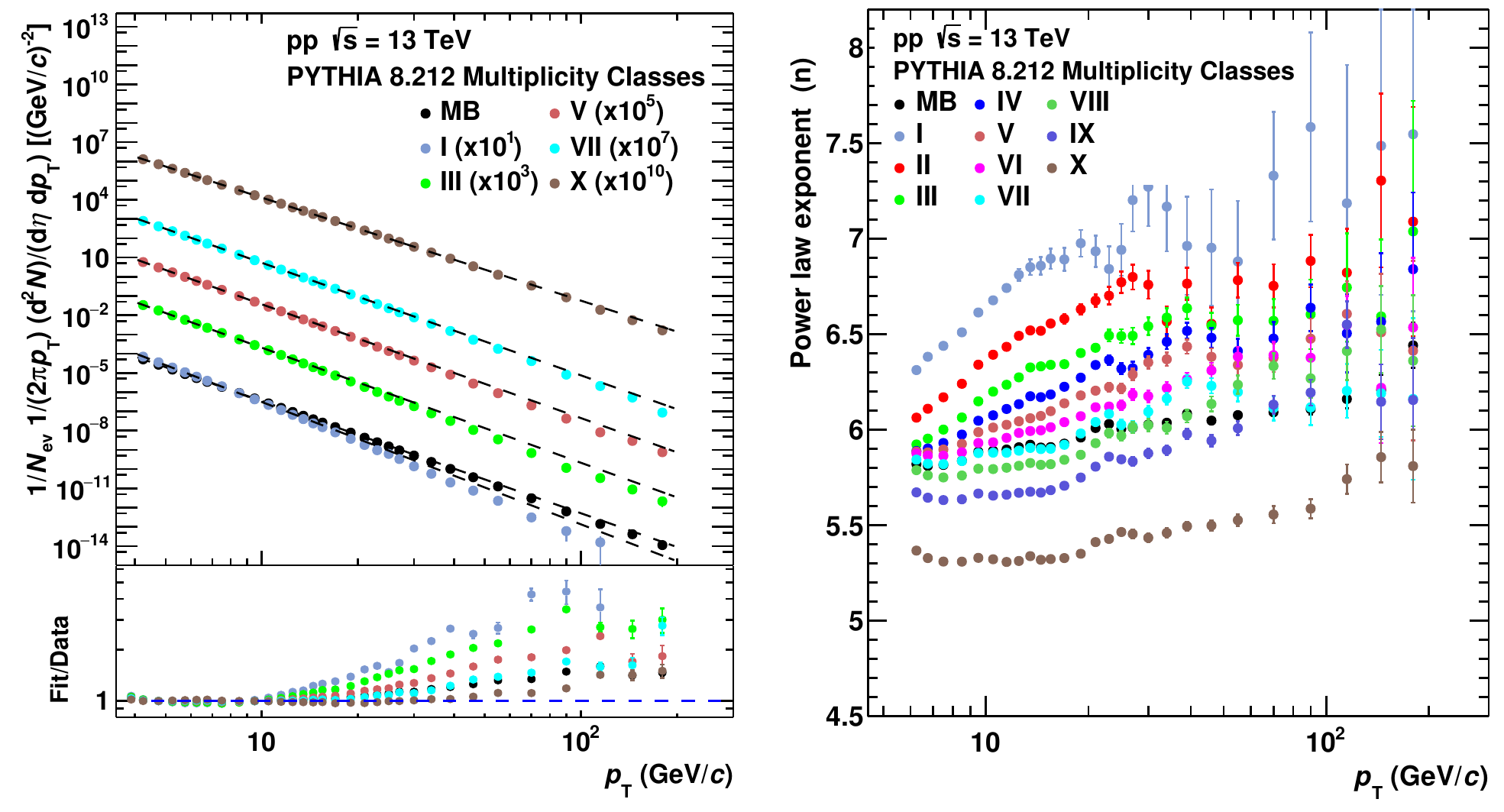}
		\caption{\label{fig:2} Left: Transverse momentum distributions of  charged particles for minimum bias (MB) and different multiplicity classes in pp collisions at $\sqrt{s}=13$\,TeV simulated with PYTHIA 8.212.  Power-law functions (dashed lines) are fitted to the \pt spectra for $\pt>8$\,GeV/$c$. Right: The power-law exponents extracted from the fits are plotted as a function of transverse momentum.} 
	\end{center}
\end{figure*}

\begin{figure*}[t!]
	\begin{center}
		\includegraphics[keepaspectratio, width=1.05\columnwidth]{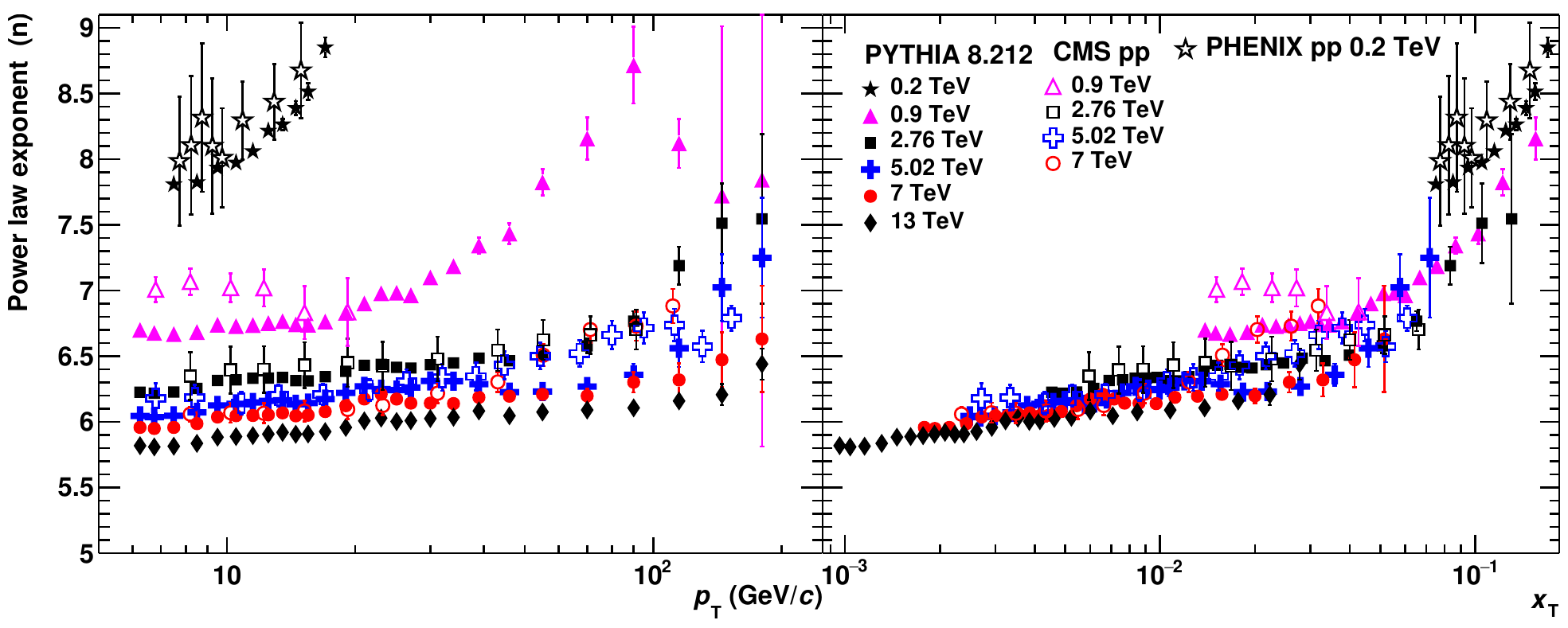}
		\caption{\label{fig:3} Power-law exponent as a function of transverse momentum (right) and $x_{\rm T}$ (left) for minimum-bias pp collisions at different energies. The data have been taken from~\cite{Khachatryan:2010xs,CMS:2012aa,Khachatryan:2016odn,Khachatryan:2010us,Abelev:2013ala,Adare:2007dg}. Results are compared with PYTHIA~8.212 predictions. } 
	\end{center}
\end{figure*}

\begin{figure*}[t!]
	\begin{center}
		\includegraphics[keepaspectratio, width=1.05\columnwidth]{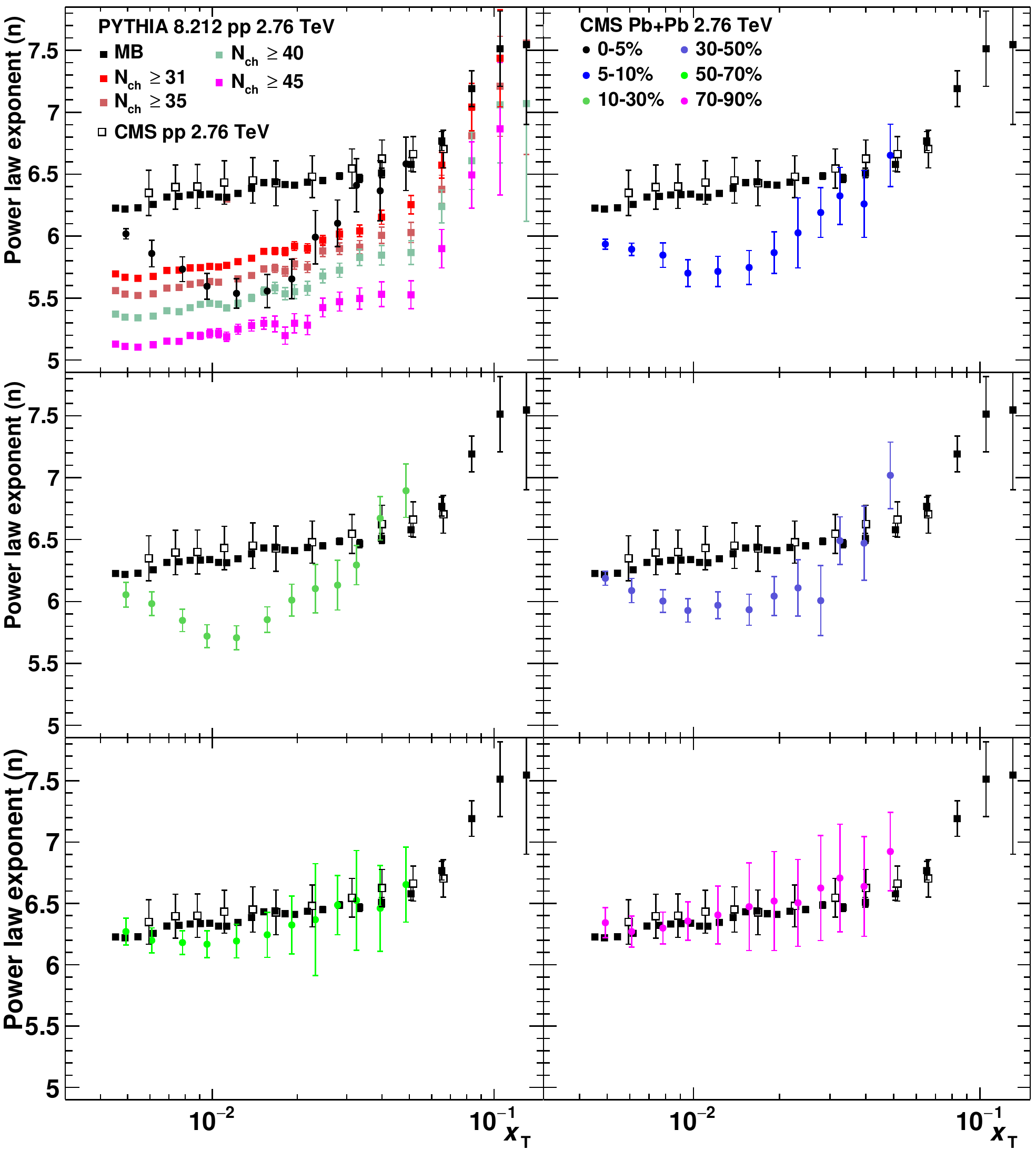}
		\caption{\label{fig:4} {Upper left: Power-law exponent as a function of $x_{\rm T}$ for central Pb-Pb collisions  at $\sqrt{s_{\rm NN}}=2.76$\,TeV. Heavy-ion data~\cite{Abelev:2012hxa} are compared with minimum-bias (MB) CMS data for pp collsions~\cite{CMS:2012aa} and high multiplicity ($|\eta|<0.8$) pp collisions at $\sqrt{s}=2.76$\,TeV simulated with PYTHIA~8.212. Other panels show the centrality dependent power-law exponents of Pb-Pb collisions as a function of $x_{\rm T}$ compared with results from MB pp collisions.} }
		
		\
		\
	\end{center}
\end{figure*}

\begin{figure*}[t!]
	\begin{center}
		\includegraphics[keepaspectratio, width=1.05\columnwidth]{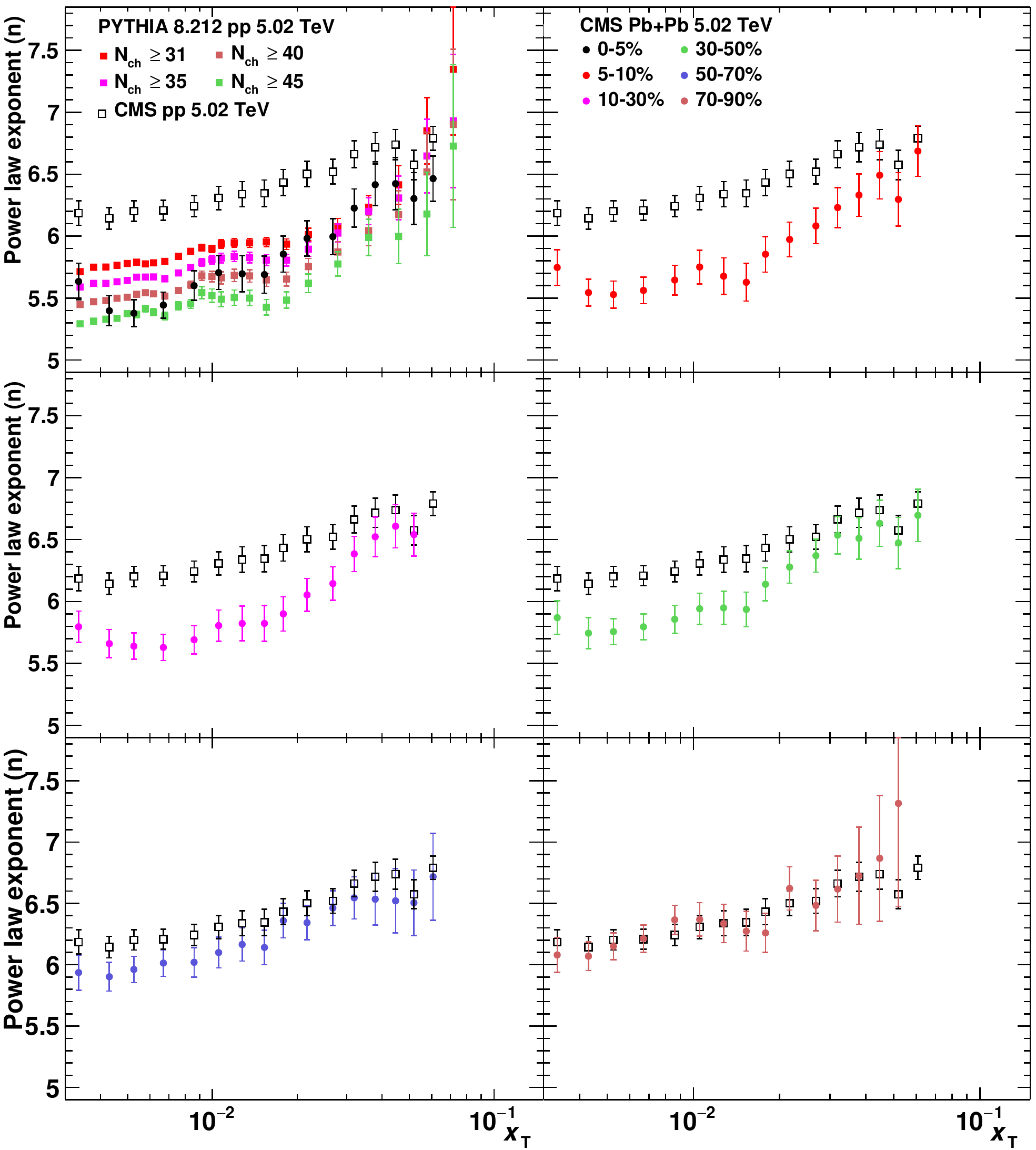}
		\caption{\label{fig:5} {Upper left: Power-law exponent as a function of $x_{\rm T}$ for central Pb-Pb collisions  at $\sqrt{s_{\rm NN}}=5.02$\,TeV. Heavy-ion data~\cite{Khachatryan:2016odn} are compared with minimum-bias and high-multiplicity ($|\eta|<0.8$) pp collisions at $\sqrt{s}=5.02$\,TeV simulated with PYTHIA~8.212. Other panels show the centrality dependent power-law exponents of Pb-Pb collisions as a function of $x_{\rm T}$ compared with results from MB pp collisions.}}
		\
		\
	\end{center}
\end{figure*}
 
\section{Results and discussion}
Figure~\ref{fig:1} shows the multiplicity dependent \pt spectra for pp collisions at $\sqrt{s}=13$\,TeV. For $\pt>8$\,GeV/$c$ the spectra become harder with increasing multiplicity. This is a consequence of the multiplicity selection bias towards hard processes which is induced when one determines the event multiplicity and the $p_{\rm T}$ spectrum within the same narrow pseudorapidity interval~\cite{Sjostrand:2014zea}. Figure~\ref{fig:1} also shows the ratios of the \pt spectra for the different multiplicity classes divided by that for minimum-bias pp collisions. They exhibit an important increase with \pt, similar to the one  observed in the \raa measured in Pb-Pb collisions~\cite{Khachatryan:2016odn}.  To characterize the changes with multiplicity we fitted a power-law function ($\propto \pt^{-n}$) to the \pt spectrum of a specific colliding system and for a given multiplicity class~\cite{Arleo:2009ch}. The power-law exponents allow us to investigate in a bias-free manner various systems, multiplicities, and energies.

Figure~\ref{fig:2} displays the multiplicity-dependent transverse momentum spectra and their corresponding fitted power-law functions for pp collisions at $\sqrt{s}=13$\,TeV. To our surprise, we observed that it is not possible to describe the full \pt (8--300\,GeV/$c$) interval assuming the same power-law exponent. For instance, for \pt larger than 20\,GeV/$c$ the ratios go beyond 20\%.  In order to check this, we have performed the fit considering sub-intervals of \pt. This allows the extraction of local power-law exponents for different \pt sub-intervals. The results indicate that the exponent has an important dependence on \pt. This is shown in the right-hand side of Fig.~\ref{fig:2} where the multiplicity dependence of $n$ as a function of \pt is shown.  The exponents have a very specific behavior with multiplicity. At low multiplicities the exponents rise more rapidly than the minimum-bias ones. At $\sqrt{s}=13$\,TeV one observes a behavior of exponents below the minimum-bias ones until quite sizable multplicities of ~ 25  are reached. Above, the spectra tend to have exponents that are smaller than observed for minimum bias. We observe that for all multiplicity classes there is a trend to have smaller exponents (softening of the spectra) at higher momenta; the tendency getting smaller for high multiplicities. Theoretically \pt can range from 0 to half of the center-of-mass energy, $\sqrt{s}/2$, of the collision. Therefore, the distribution can also be presented as a function of the dimensionless variable $x_{\rm T}=2\pt/\sqrt{s}$~\cite{SIVERS19761}, which varies between 0 and 1. 

In Fig.~\ref{fig:3} we show $n$ as a function of $p_{\rm T}$ for minimum-bias pp data at different $\sqrt{s}$ (0.2, 0.9, 2.76, 5.02, 7 and 13\,TeV~\cite{Khachatryan:2010xs,CMS:2012aa,Khachatryan:2016odn,Khachatryan:2010us,Abelev:2013ala,Adare:2007dg}). The results are compared with PYTHIA~8.212~\cite{Sjostrand:2014zea} (tune Monash~2013~\cite{Skands:2014pea}) simulations. Going from low to high energies the power-law exponent decreases in both data and PYTHIA~8.212. This is expected since at higher energies the production cross sections of hard processes increase resulting in a change in the slope of the spectra at large transverse momenta. A different representation is shown in the right-hand-side plot, where the power-law exponent is presented as a function of $x_{\rm T}$.  Within 10\% the data, that were before distinctly different, fall now approximately on an universal curve. Prominent in this respect is the case of the $\sqrt{s}=0.2$\,TeV data. The approximate scaling property is well reproduced by PYTHIA 8.212.

Applying now the same treatment to the Pb-Pb data we observe that the exponents as a function of $x_{\rm T}$ and centrality behave very similar to those for pp collisions simulated with PYTHIA 8.212. The comparison in shown in  Fig.~\ref{fig:4} for pp and Pb-Pb collisions at $\sqrt{s_{\rm NN}}=2.76$\,TeV. Going from  70--90\% peripheral to 0--5\% central Pb-Pb collisions the exponent exhibits an overall decrease for $x_{\rm T}$ below 0.02. This is consistent with the hardening of the $p_{\rm T}$ spectra going from peripheral to central Pb-Pb collisions. For higher $x_{\rm T}$ (\textgreater 0.02) the exponents gradually rise towards the minimum-bias value at $x_{\rm T}\approx0.04$. This behavior is consistent with the approaching to unity of  $R_{\rm AA}$ at very high $p_{\rm T}$~\cite{Loizides:2014vua}.  The multiplicity dependence of $n$ vs $x_{\rm T}$ in pp collisions simulated with PYTHIA 8.212 is qualitatively similar to that observed in heavy-ion data. The same behavior is also observed at higher energies; in particular Fig.~\ref{fig:5} shows results for Pb-Pb collisions at $\sqrt{s_{\rm NN}}=5.02$\,TeV. The observed behavior invites one interesting consequence. The accepted view which entirely attributes the rise in \raa to the decrease of the parton energy loss should be revised. It is well known that the mean \pt continues rising  with multiplicity both in pp and in heavy-ion collisions, implying that high multiplicity, which is proportional to the energy density, is correlated with the high momentum particle production.

\section{Conclusion}

We have studied the high-\pt ($\pt>8$\,GeV/$c$) charged-particle production in both pp and Pb-Pb collisions. Considering different \pt subintervals, power-law functions were fitted to the transverse momentum distributions of minimum-bias pp collisions measured by experiments at the RHIC and LHC. The local exponents of the power-law fits were compared to those obtained from Pb-Pb data. Using PYTHIA~8 simulations, we also studied the charged-particle multiplicity dependence of the exponent in pp collisions. With respect to minimum-bias pp collisions, we have determined the following:   

\begin{itemize}
	\item The high-\pt part of the \pt spectra cannot be described by a single power-law function (same exponent value) within a wide \pt interval (8-100\,GeV/$c$). 
	\item The minimum-bias \pt spectra, when represented in terms of the local exponent as a function of the Bjorken variable $x_{\rm T}$, obey an approximate scaling behavior over a wide range of center-of-mass energy, $\sqrt{s}=0.2$ to 13\,TeV.
	\item The \pt spectral shape (characterized by  local exponents) as a function of multiplicity exhibits a specific behavior. For $8<\pt<30$\,GeV/$c$ the local exponents are smaller than those for minimum-bias events, i.e. the \pt spectra are harder for high-multiplicity events than that for minimum-bias pp collisions. At higher \pt (30-100\,GeV/$c$) the exponents gradually increase to reach the values which describe the minimum-bias \pt spectra.
	\item For heavy-ion collisions the evolution of the local exponent as a function of $x_{T}$ and collision centrality is qualitatively similar to that for pp collisions. The only specific difference is that the heavy-ion data show a particular shape of the exponent evolution with a downward trend for lower values of $x_{\rm T}$ (\pt) . This is not observed in pp collision, but one has to consider that PYTHIA~8 does not necessarily describe the multiplicity-dependent pp data. Unfortunately at the present, pp data for different multiplicity classes and wide \pt intervals  are not available.
\end{itemize}

It would be very important to produce experimental results on high-multiplicity pp collisions over a wide \pt interval in order to be able to assess in details the source of the apparent similarity between pp and A-A data. 



\section{Acknowledgments}
Support for this work has been received from CONACyT under the grant number 280362 and PAPIIT-UNAM under Project No. IN102118. G.P. thanks the DGAPA, the Centro Fermi and the ALICE collaboration for their support. A.M. acknowledges the postdoctoral fellowship of DGAPA UNAM.

\section*{References}

\bibliography{MyRefFile}

\begin{thebibliography}{10}
\expandafter\ifx\csname url\endcsname\relax
  \def\url#1{\texttt{#1}}\fi
\expandafter\ifx\csname urlprefix\endcsname\relax\def\urlprefix{URL }\fi
\expandafter\ifx\csname href\endcsname\relax
  \def\href#1#2{#2} \def\path#1{#1}\fi

\bibitem{edward:2010aip}
E.~K.~G. Sarkisyan, A.~S. Sakharov, {Multihadron production features in
  different reactions}, AIP Conf. Proc. 828 (2006) 35.
\newblock \href {http://arxiv.org/abs/0510191} {\path{arXiv:0510191}}, \href
  {http://dx.doi.org/10.1063/1.2197392} {\path{doi:10.1063/1.2197392}}.

\bibitem{edward:2010epjc}
E.~K.~G. Sarkisyan, A.~S. Sakharov, {Relating multihadron production in
  hadronic and nuclear collisions}, Eur. Phys. J. C70 (2010) 533--541.
\newblock \href {http://arxiv.org/abs/1004.4390} {\path{arXiv:1004.4390}},
  \href {http://dx.doi.org/10.1140/epjc/s10052-010-1493-1}
  {\path{doi:10.1140/epjc/s10052-010-1493-1}}.

\bibitem{edward:2014epjc}
A.~N. Mishra, R.~Sahoo, E.~K.~G. Sarkisyan, A.~S. Sakharov, {Effective-energy
  budget in multiparticle production in nuclear collisions}, Eur. Phys. J. C74
  (2014) 3147.
\newblock \href {http://arxiv.org/abs/1405.2819} {\path{arXiv:1405.2819}},
  \href {http://dx.doi.org/10.1140/epjc/s10052-015-3275-2}
  {\path{doi:10.1140/epjc/s10052-015-3275-2}}.

\bibitem{edward:2016prd}
E.~K.~G. Sarkisyan, A.~N. Mishra, R.~Sahoo, A.~S. Sakharov, {Multihadron
  production dynamics exploring the energy balance in hadronic and nuclear
  collisions}, Phys. Rev. D93 (2016) 054046.
\newblock \href {http://arxiv.org/abs/1506.09080} {\path{arXiv:1506.09080}},
  \href {http://dx.doi.org/10.1103/PhysRevD.93.054046}
  {\path{doi:10.1103/PhysRevD.93.054046}}.

\bibitem{edward:2016Rprd}
E.~K.~G. Sarkisyan, A.~N. Mishra, R.~Sahoo, A.~S. Sakharov, {Centrality
  dependence of midrapidity density from GeV to TeV heavy-ion collisions in the
  effective-energy universality picture of hadroproduction}, Phys. Rev. D94
  (2016) 011501.
\newblock \href {http://arxiv.org/abs/1603.09040} {\path{arXiv:1603.09040}},
  \href {http://dx.doi.org/10.1103/PhysRevD.94.011501}
  {\path{doi:10.1103/PhysRevD.94.011501}}.

\bibitem{Loizides:2016tew}
C.~Loizides, {Experimental overview on small collision systems at the LHC},
  Nucl. Phys. A956 (2016) 200--207.
\newblock \href {http://arxiv.org/abs/1602.09138} {\path{arXiv:1602.09138}},
  \href {http://dx.doi.org/10.1016/j.nuclphysa.2016.04.022}
  {\path{doi:10.1016/j.nuclphysa.2016.04.022}}.

\bibitem{Abelev:2013haa}
B.~B. Abelev, et~al., {Multiplicity Dependence of Pion, Kaon, Proton and Lambda
  Production in p-Pb Collisions at $\sqrt{s_{NN}}$ = 5.02 TeV}, Phys. Lett.
  B728 (2014) 25--38.
\newblock \href {http://arxiv.org/abs/1307.6796} {\path{arXiv:1307.6796}},
  \href {http://dx.doi.org/10.1016/j.physletb.2013.11.020}
  {\path{doi:10.1016/j.physletb.2013.11.020}}.

\bibitem{Khachatryan:2016txc}
V.~Khachatryan, et~al., {Evidence for collectivity in pp collisions at the
  LHC}, Phys. Lett. B765 (2017) 193--220.
\newblock \href {http://arxiv.org/abs/1606.06198} {\path{arXiv:1606.06198}},
  \href {http://dx.doi.org/10.1016/j.physletb.2016.12.009}
  {\path{doi:10.1016/j.physletb.2016.12.009}}.

\bibitem{ALICE:2017jyt}
J.~Adam, et~al., {Enhanced production of multi-strange hadrons in
  high-multiplicity proton-proton collisions}, Nature Phys. 13 (2017) 535--539.
\newblock \href {http://arxiv.org/abs/1606.07424} {\path{arXiv:1606.07424}},
  \href {http://dx.doi.org/10.1038/nphys4111} {\path{doi:10.1038/nphys4111}}.

\bibitem{Zakharov:2013gya}
B.~G. Zakharov, {Parton energy loss in the mini quark-gluon plasma and jet
  quenching in proton-proton collisions}, J. Phys. G41 (2014) 075008.
\newblock \href {http://arxiv.org/abs/1311.1159} {\path{arXiv:1311.1159}},
  \href {http://dx.doi.org/10.1088/0954-3899/41/7/075008}
  {\path{doi:10.1088/0954-3899/41/7/075008}}.

\bibitem{Mangano:2017plv}
M.~L. Mangano, B.~Nachman, {Observables for possible QGP signatures in central
  pp collisions}\href {http://arxiv.org/abs/1708.08369}
  {\path{arXiv:1708.08369}}.

\bibitem{Adare:2015cua}
A.~Adare, et~al., {Scaling properties of fractional momentum loss of high-$p_T$
  hadrons in nucleus-nucleus collisions at $\sqrt{s_{_{NN}}}$ from 62.4 GeV to
  2.76 TeV}, Phys. Rev. C93~(2) (2016) 024911.
\newblock \href {http://arxiv.org/abs/1509.06735} {\path{arXiv:1509.06735}},
  \href {http://dx.doi.org/10.1103/PhysRevC.93.024911}
  {\path{doi:10.1103/PhysRevC.93.024911}}.

\bibitem{Ortiz:2017cul}
A.~Ortiz, O.~V\'azquez, {Energy density and path-length dependence of the
  fractional momentum loss in heavy-ion collisions at $\sqrt{s_{\rm NN}}$ from
  62.4 to 5020 GeV}, Phys. Rev. C97~(1) (2018) 014910.
\newblock \href {http://arxiv.org/abs/1708.07571} {\path{arXiv:1708.07571}},
  \href {http://dx.doi.org/10.1103/PhysRevC.97.014910}
  {\path{doi:10.1103/PhysRevC.97.014910}}.

\bibitem{Acharya:2018qsh}
S.~Acharya, et~al., {Transverse momentum spectra and nuclear modification
  factors of charged particles in pp, p-Pb and Pb-Pb collisions at the
  LHC}\href {http://arxiv.org/abs/1802.09145} {\path{arXiv:1802.09145}}.

\bibitem{Acharya:2018eaq}
S.~Acharya, et~al., {Transverse momentum spectra and nuclear modification
  factors of charged particles in Xe-Xe collisions at $\sqrt{s_{\rm NN}}$ =
  5.44 TeV}\href {http://arxiv.org/abs/1805.04399} {\path{arXiv:1805.04399}}.

\bibitem{ALICE:2011ac}
B.~Abelev, et~al., {Underlying Event measurements in $pp$ collisions at
  $\sqrt{s}=0.9$ and 7 TeV with the ALICE experiment at the LHC}, JHEP 07
  (2012) 116.
\newblock \href {http://arxiv.org/abs/1112.2082} {\path{arXiv:1112.2082}},
  \href {http://dx.doi.org/10.1007/JHEP07(2012)116}
  {\path{doi:10.1007/JHEP07(2012)116}}.

\bibitem{Ortiz:2017jaz}
A.~Ortiz, L.~Valencia~Palomo, {Universality of the underlying event in pp
  collisions}, Phys. Rev. D96~(11) (2017) 114019.
\newblock \href {http://arxiv.org/abs/1710.04741} {\path{arXiv:1710.04741}},
  \href {http://dx.doi.org/10.1103/PhysRevD.96.114019}
  {\path{doi:10.1103/PhysRevD.96.114019}}.

\bibitem{Armesto:2015kwa}
N.~Armesto, D.~C. Gülhan, J.~G. Milhano, {Kinematic bias on centrality
  selection of jet events in pPb collisions at the LHC}, Phys. Lett. B747
  (2015) 441--445.
\newblock \href {http://arxiv.org/abs/1502.02986} {\path{arXiv:1502.02986}},
  \href {http://dx.doi.org/10.1016/j.physletb.2015.06.032}
  {\path{doi:10.1016/j.physletb.2015.06.032}}.

\bibitem{Alvioli:2014eda}
M.~Alvioli, B.~A. Cole, L.~Frankfurt, D.~V. Perepelitsa, M.~Strikman, {Evidence
  for $x$-dependent proton color fluctuations in pA collisions at the CERN
  Large Hadron Collider}, Phys. Rev. C93~(1) (2016) 011902.
\newblock \href {http://arxiv.org/abs/1409.7381} {\path{arXiv:1409.7381}},
  \href {http://dx.doi.org/10.1103/PhysRevC.93.011902}
  {\path{doi:10.1103/PhysRevC.93.011902}}.

\bibitem{Ortiz:2018vgc}
A.~Ortiz, L.~Valencia~Palomo, {Probing color reconnection with underlying event
  observables at the LHC energies}, Phys. Rev. D99~(3) (2019) 034027.
\newblock \href {http://arxiv.org/abs/1809.01744} {\path{arXiv:1809.01744}},
  \href {http://dx.doi.org/10.1103/PhysRevD.99.034027}
  {\path{doi:10.1103/PhysRevD.99.034027}}.

\bibitem{Horowitz:2011gd}
W.~A. Horowitz, M.~Gyulassy, {The Surprising Transparency of the sQGP at LHC},
  Nucl. Phys. A872 (2011) 265--285.
\newblock \href {http://arxiv.org/abs/1104.4958} {\path{arXiv:1104.4958}},
  \href {http://dx.doi.org/10.1016/j.nuclphysa.2011.09.018}
  {\path{doi:10.1016/j.nuclphysa.2011.09.018}}.

\bibitem{Alver:2008aq}
B.~Alver, M.~Baker, C.~Loizides, P.~Steinberg, {The PHOBOS Glauber Monte
  Carlo}\href {http://arxiv.org/abs/0805.4411} {\path{arXiv:0805.4411}}.

\bibitem{Loizides:2014vua}
C.~Loizides, J.~Nagle, P.~Steinberg, {Improved version of the PHOBOS Glauber
  Monte Carlo}, SoftwareX 1-2 (2015) 13--18.
\newblock \href {http://arxiv.org/abs/1408.2549} {\path{arXiv:1408.2549}},
  \href {http://dx.doi.org/10.1016/j.softx.2015.05.001}
  {\path{doi:10.1016/j.softx.2015.05.001}}.

\bibitem{Khachatryan:2016odn}
V.~Khachatryan, et~al., {Charged-particle nuclear modification factors in PbPb
  and pPb collisions at $ \sqrt{s_{\mathrm{N}\;\mathrm{N}}}=5.02 $ TeV}, JHEP
  04 (2017) 039.
\newblock \href {http://arxiv.org/abs/1611.01664} {\path{arXiv:1611.01664}},
  \href {http://dx.doi.org/10.1007/JHEP04(2017)039}
  {\path{doi:10.1007/JHEP04(2017)039}}.

\bibitem{Abelev:2014laa}
B.~B. Abelev, et~al., {Production of charged pions, kaons and protons at large
  transverse momenta in pp and PbPb collisions at $\sqrt{s_{\rm NN}}$ =2.76
  TeV}, Phys. Lett. B736 (2014) 196--207.
\newblock \href {http://arxiv.org/abs/1401.1250} {\path{arXiv:1401.1250}},
  \href {http://dx.doi.org/10.1016/j.physletb.2014.07.011}
  {\path{doi:10.1016/j.physletb.2014.07.011}}.

\bibitem{Adam:2015kca}
J.~Adam, et~al., {Centrality dependence of the nuclear modification factor of
  charged pions, kaons, and protons in Pb-Pb collisions at $\sqrt{s_{\rm
  NN}}=2.76$ TeV}, Phys. Rev. C93~(3) (2016) 034913.
\newblock \href {http://arxiv.org/abs/1506.07287} {\path{arXiv:1506.07287}},
  \href {http://dx.doi.org/10.1103/PhysRevC.93.034913}
  {\path{doi:10.1103/PhysRevC.93.034913}}.

\bibitem{Abelev:2012sk}
B.~Abelev, et~al., {Transverse sphericity of primary charged particles in
  minimum bias proton-proton collisions at $\sqrt{s}=0.9$, 2.76 and 7 TeV},
  Eur. Phys. J. C72 (2012) 2124.
\newblock \href {http://arxiv.org/abs/1205.3963} {\path{arXiv:1205.3963}},
  \href {http://dx.doi.org/10.1140/epjc/s10052-012-2124-9}
  {\path{doi:10.1140/epjc/s10052-012-2124-9}}.

\bibitem{CMS:2016gme}
S.~Chatrchyan, et~al., {Dependence of the $\Upsilon$(nS) production ratios on
  charged particle multiplicity in pp collisions at $\sqrt{s}=7~\mathrm{TeV}$}
  (2016).

\bibitem{Sjostrand:2014zea}
T.~Sj{\"o}strand, S.~Ask, J.~R. Christiansen, R.~Corke, N.~Desai, P.~Ilten,
  S.~Mrenna, S.~Prestel, C.~O. Rasmussen, P.~Z. Skands, {An Introduction to
  PYTHIA 8.2}, Comput. Phys. Commun. 191 (2015) 159--177.
\newblock \href {http://arxiv.org/abs/1410.3012} {\path{arXiv:1410.3012}},
  \href {http://dx.doi.org/10.1016/j.cpc.2015.01.024}
  {\path{doi:10.1016/j.cpc.2015.01.024}}.

\bibitem{Ortiz:2016kpz}
A.~Ortiz, G.~Bencedi, H.~Bello, {Revealing the source of the radial flow
  patterns in proton-proton collisions using hard probes}, J. Phys. G44~(6)
  (2017) 065001.
\newblock \href {http://arxiv.org/abs/1608.04784} {\path{arXiv:1608.04784}},
  \href {http://dx.doi.org/10.1088/1361-6471/aa6594}
  {\path{doi:10.1088/1361-6471/aa6594}}.

\bibitem{Skands:2014pea}
P.~Skands, S.~Carrazza, J.~Rojo, {Tuning PYTHIA 8.1: the Monash 2013 Tune},
  Eur. Phys. J. C74~(8) (2014) 3024.
\newblock \href {http://arxiv.org/abs/1404.5630} {\path{arXiv:1404.5630}},
  \href {http://dx.doi.org/10.1140/epjc/s10052-014-3024-y}
  {\path{doi:10.1140/epjc/s10052-014-3024-y}}.

\bibitem{Adam:2015pza}
J.~Adam, et~al., {Pseudorapidity and transverse-momentum distributions of
  charged particles in proton-proton collisions at $\sqrt s=$ 13 TeV}, Phys.
  Lett. B753 (2016) 319--329.
\newblock \href {http://arxiv.org/abs/1509.08734} {\path{arXiv:1509.08734}},
  \href {http://dx.doi.org/10.1016/j.physletb.2015.12.030}
  {\path{doi:10.1016/j.physletb.2015.12.030}}.

\bibitem{Ortiz:2013yxa}
A.~Ortiz, P.~Christiansen, E.~Cuautle, I.~Maldonado, G.~Pai{\'c}, {Color
  Reconnection and Flowlike Patterns in $pp$ Collisions}, Phys. Rev. Lett.
  111~(4) (2013) 042001.
\newblock \href {http://arxiv.org/abs/1303.6326} {\path{arXiv:1303.6326}},
  \href {http://dx.doi.org/10.1103/PhysRevLett.111.042001}
  {\path{doi:10.1103/PhysRevLett.111.042001}}.

\bibitem{Khachatryan:2010xs}
V.~Khachatryan, et~al., {Transverse momentum and pseudorapidity distributions
  of charged hadrons in pp collisions at $\sqrt{s} = 0.9$ and 2.36 TeV}, JHEP
  02 (2010) 041.
\newblock \href {http://arxiv.org/abs/1002.0621} {\path{arXiv:1002.0621}},
  \href {http://dx.doi.org/10.1007/JHEP02(2010)041}
  {\path{doi:10.1007/JHEP02(2010)041}}.

\bibitem{CMS:2012aa}
S.~Chatrchyan, et~al., {Study of high-pT charged particle suppression in PbPb
  compared to $pp$ collisions at $\sqrt{s_{NN}}=2.76$ TeV}, Eur. Phys. J. C72
  (2012) 1945.
\newblock \href {http://arxiv.org/abs/1202.2554} {\path{arXiv:1202.2554}},
  \href {http://dx.doi.org/10.1140/epjc/s10052-012-1945-x}
  {\path{doi:10.1140/epjc/s10052-012-1945-x}}.

\bibitem{Khachatryan:2010us}
V.~Khachatryan, et~al., {Transverse-momentum and pseudorapidity distributions
  of charged hadrons in $pp$ collisions at $\sqrt{s}=7$ TeV}, Phys. Rev. Lett.
  105 (2010) 022002.
\newblock \href {http://arxiv.org/abs/1005.3299} {\path{arXiv:1005.3299}},
  \href {http://dx.doi.org/10.1103/PhysRevLett.105.022002}
  {\path{doi:10.1103/PhysRevLett.105.022002}}.

\bibitem{Abelev:2013ala}
B.~B. Abelev, et~al., {Energy Dependence of the Transverse Momentum
  Distributions of Charged Particles in pp Collisions Measured by ALICE}, Eur.
  Phys. J. C73~(12) (2013) 2662.
\newblock \href {http://arxiv.org/abs/1307.1093} {\path{arXiv:1307.1093}},
  \href {http://dx.doi.org/10.1140/epjc/s10052-013-2662-9}
  {\path{doi:10.1140/epjc/s10052-013-2662-9}}.

\bibitem{Adare:2007dg}
A.~Adare, et~al., {Inclusive cross-section and double helicity asymmetry for
  pi0 production in p + p collisions at s**(1/2) = 200-GeV: Implications for
  the polarized gluon distribution in the proton}, Phys. Rev. D76 (2007)
  051106.
\newblock \href {http://arxiv.org/abs/0704.3599} {\path{arXiv:0704.3599}},
  \href {http://dx.doi.org/10.1103/PhysRevD.76.051106}
  {\path{doi:10.1103/PhysRevD.76.051106}}.

\bibitem{Abelev:2012hxa}
B.~Abelev, et~al., {Centrality Dependence of Charged Particle Production at
  Large Transverse Momentum in Pb--Pb Collisions at $\sqrt{s_{\rm{NN}}} = 2.76$
  TeV}, Phys. Lett. B720 (2013) 52--62.
\newblock \href {http://arxiv.org/abs/1208.2711} {\path{arXiv:1208.2711}},
  \href {http://dx.doi.org/10.1016/j.physletb.2013.01.051}
  {\path{doi:10.1016/j.physletb.2013.01.051}}.

\bibitem{Arleo:2009ch}
F.~Arleo, S.~J. Brodsky, D.~S. Hwang, A.~M. Sickles, {Higher-Twist Dynamics in
  Large Transverse Momentum Hadron Production}, Phys. Rev. Lett. 105 (2010)
  062002.
\newblock \href {http://arxiv.org/abs/0911.4604} {\path{arXiv:0911.4604}},
  \href {http://dx.doi.org/10.1103/PhysRevLett.105.062002}
  {\path{doi:10.1103/PhysRevLett.105.062002}}.

\bibitem{SIVERS19761}
D.~Sivers, S.~J. Brodsky, R.~Blankenbecler,
  \href{http://www.sciencedirect.com/science/article/pii/0370157376900156}{Large
  transverse momentum processes}, Physics Reports 23~(1) (1976) 1 -- 121.
\newblock \href
  {http://dx.doi.org/https://doi.org/10.1016/0370-1573(76)90015-6}
  {\path{doi:https://doi.org/10.1016/0370-1573(76)90015-6}}.
\newline\urlprefix\url{http://www.sciencedirect.com/science/article/pii/0370157376900156}

\end{thebibliography}

\end{document}